\def\la{\lesssim} 
\def\ga{\hbox{{\lower -2.5pt\hbox{$>$}}\hskip -8pt\raise
-2.5pt\hbox{$\sim$}}}

\documentstyle[preprint,aps]{revtex}
\tightenlines
\begin{document}

\title {A Magnetized Local Supercluster and the Origin of
the Highest Energy Cosmic Rays}
\author{Pasquale Blasi
 \footnote{e-mail: blasi@oddjob.uchicago.edu}
 and Angela V. Olinto
 \footnote{e-mail: olinto@oddjob.uchicago.edu}}

\address{Department of Astronomy \& Astrophysics, and \\
Enrico Fermi Institute, The University of Chicago,\\
5640 S. Ellis Av., Chicago, IL 60637}
\maketitle
\begin{abstract}
A sufficiently magnetized Local Supercluster can explain the spectrum and
angular distribution of ultra-high energy cosmic rays. We show  that the
spectrum of extragalactic cosmic rays with energies below $\sim 10^{20}$
eV may be due to the diffusive propagation in the Local Supercluster with
fields of  $\sim 10^{-8} - 10^{-7}$ Gauss. Above  $\sim 10^{20}$ eV,
cosmic rays propagate in an almost rectilinear way which is evidenced by
the change in shape of the spectrum at the highest energies.  The fit to
the spectrum requires that at least one  source be located relatively
nearby at $\sim  10-15$ Mpc away from the Milky Way. We discuss the origin
of magnetic fields in the Local Supercluster and the observable
predictions of this model.
\vskip 0.3cm
\noindent PACS numbers: 98.70.Sa, 96.40.De, 95.85.Sz, 98.38.Am, 98.62.-g   
\vskip 0.3 cm
\centerline{\it Submitted to Physical Review D}

\end{abstract}

\section{Introduction}

The observed cosmic ray spectrum covers about 11 orders of
magnitude, from about a GeV to a few $10^{11}$ GeV, with the remarkable
property of being almost featureless. Up to a ``knee,'' at $\sim 10^{15}$
eV, the power law spectrum is believed to be  due to shock acceleration
in supernova remnants\cite{LSC}.  Above the knee, the spectrum changes to
a slightly steeper power law down to an ``ankle'' at $\sim 10^{19}$ eV
where a flattening is observed.  Between the knee and the ankle, cosmic
rays are considered to be produced in the Galaxy, though the mechanism is
not yet clear. Above
$10^{19}$ eV, the proton gyroradius in the Galactic  magnetic field is
larger than the size of the halo. This together with the lack of
identified Galactic sources in the arrival direction of the events
suggest an extragalactic origin for these ultra-high energy cosmic rays
(UHECRs).

If the sources of UHECRs are extragalactic, then their distances cannot be
much larger than $50-100$ Mpc due to the photopion production off the
cosmic background radiation\cite{GZK}. Protons with energies above  $\sim
5\times 10^{19}$ eV experience significant losses  manifested as the
Greisen-Zatsepin-Kuzmin (GZK) cutoff in the spectrum. This limitation in
the volume available for sources together with the extreme energies
required for acceleration and the lack of source counterpart
identifications make the origin of the highest energy events a
challenging mystery.

Crucial for the ability to point back at possible sources is the magnetic
structure of the medium between the source and us. At present, the
structure of magnetic fields in the extragalactic medium is poorly known.
Evidence for  equipartition fields in some clusters of galaxies is well
established while extragalactic fields on very large scales are
constrained by Faraday rotation measures from distant quasars to
magnitudes below
$\sim 10^{-9}$ Gauss for a reversal scale of 1 Mpc\cite{K95}.  The
deflection angle of UHECRs in extragalactic fields below this upper limit
is only a few degrees for sources closer than
$100$ Mpc, therefore the sources should be {\it visible} in other
wavelenghts within the error box of the detected events. To date, no
plausible source counterparts have been found within a few degrees from the
arrival direction.

Here we argue that the lack of counterpart identifications for energies
below $10^{20}$ eV is a natural consequence of a close to diffusive
propagation in a magnetized Local Supercluster (LSC) and that counterparts
may be more easily identified at energies above the limit of
detectability of current experiments. In addition, we show that the
transition from diffusive to straight line propagation of UHECRs in the
LSC gives a good fit to the spectral changes observed with current
experiments. These findings show that a simple model for UHECRs based on
one or more nearby continuos sources can explain the presently available
data on UHECRs if the LSC has relatively strong magnetic fields ($\sim
10^{-8}-10^{-7}$ Gauss). Such fields are allowed by current limits on
extragalactic magnetic fields and could indicate the presence of magnetic
fields on cosmological scales.

\section{The Local Supercluster and Its Magnetic Field}

The Local Supercluster is a flattened overdensity of galaxies extending
for $\sim 30 - 40 $ Mpc with a width of about $\sim 10$ Mpc\cite{LSC}. Our
Galaxy is located at the edge of this distribution, while the  Virgo
cluster is approximately in the center,  $\sim 17$ Mpc away from us. This
large scale distribution  intersects almost  perpendicularly the Galactic
plane defining  the Supergalactic Plane and is most prominent in the
northern Galactic hemisphere.

Sources of UHECRs are likely to be embedded in the LSC since it contains
several active galaxies which can produce energetic particles and are
close enough to survive the photopion energy losses. Thus, it would be
natural to assume a correlation between the arrival direction of UHECRs
and the location of the supergalactic plane. Early studies suggested a
significant correlation  for cosmic rays with energies $E > 4 \times
10^{19}$ eV \cite{Stanev}, but further analysis using the larger AGASA
data set did not find appreciable departures from an isotropic
distribution in the observable region of the sky \cite{AGASA}. In the
scenario proposed here, the correlations should only  become significant
for energies $E > 10^{20}$ eV.

The magnetic field structure of the LSC is very poorly known. Direct
measurements through Faraday rotation give  upper limits to fields
averaged along the line of sight to particular background radio sources.
Assuming a reversal scale of 1 Mpc,  magnetic fields present over
cosmological distances are constrained to have ${\bar B}_{\rm Mpc} \la 
10^{-9}$ Gauss, while for reversal scales close to the present horizon,
${\bar B}_{\rm H_0^{-1}} \la 10 ^{-11}$ Gauss, if most of the Universe is
filled with plasma. If there are significant voids in the plasma
distribution in the Universe, ${\bar B}_{\rm Mpc}$ may exceed the
$10^{-9}$ Gauss limit. In addition, fields with smaller reversal scales
or in denser regions of the Universe can be significantly stronger than
these limiting strengths. In fact, direct observations of a bridge-like
structure  that connects A1367 and the Coma clusters, shows the presence
of magnetic fields of the order of $\mu G$ far from the clusters'
core\cite{ComaB}. The LSC is likely to have similar overdensities to
these filamentary structures and  have fields that greatly exceed the
$10^{-9}$ Gauss limit. In fact, Vallee \cite{Vallee} showed that in the
region within $20^o$ around the center of the Virgo cluster, Faraday
rotation measures indicate fields of about 1.5$\times 10^{-6}$Gauss.

Superclusters are most probably not virialized structures. Therefore,
equipartition magnetic fields are unlikely to be generated  by the
gravitational dynamics of these large scale systems alone. Extragalactic
fields such as those observed around the Coma cluster are more likely to
originate from local pollution by normal and radio galaxies  or through
the amplification of a primordial magnetic field.  These two
possibilities imply different structures for the  present  large scale
magnetic fields. In particular, the structure on the largest scales and
the filling factors for  extragalactic fields are likely to depend on the
origin of the fields.

Magnetized outflows from  galaxies such as jets and winds efficiently
pollute the extragalactic medium if most galaxies lived through an active
phase in their history
\cite{KL}. If intercluster fields are due to galactic outflows, the
resulting field will be randomly oriented within cells of sizes below the
mean separation between galaxies, i.e.,
$l_{cell} \la  0.5-1$ Mpc. The outflows are likely to generate turbulent
fields with structure on scales below $l_{cell}$.

Primordial magnetic fields can also give rise to large scale magnetic
fields. As superclusters form, primordial fields are amplified due to the
density increase ($B \propto \rho^{2/3}$) and due to peculiar velocities
of the gravitating gas. Overdensities in superclusters vary between 1 to
100 times the critical density, therefore supercluster fields may reach
$\sim 20$ times the average cosmological magnetic field on Mpc scales as
the LSC starts to form. In addition, peculiar motions  within the LSC may
amplify a seed primordial field to $\la 10^{-6}$ Gauss \cite{Bierm}. 

Although primordial fields start with very little structure below 10 Mpc
after recombination
\cite{JKO}, peculiar motions induced by gravitational collapse generate
magnetic modes on smaller scales. Numerical simulations of structure
formation indicate the development of a cascade to high wavenumbers
similar to a Kolmogorov spectrum \cite{CROK}.  Therefore, the spectrum of
magnetic fields generated by primordial fields or by outflows may not be
easily distinguished on scales below the mean separation between galaxies.

As we discuss below, the energy dependence of the diffusion
coefficient for UHECRs depends primarily on the strength and spectrum of
the magnetic field in the LSC. Given the present observations,  we
consider the   strength of the average magnetic field in the LSC to lie
in the range between $10^{-8}-10^{-7}$ Gauss  with a Kolmogorov spectrum
below the largest eddy scale of $l_{cell} \la \, 0.5-1$ Mpc.

\section{The propagation of Cosmic Rays in the Local Supercluster}

Most recent studies of the propagation of UHECR in the intergalactic
medium are based on the assumption that the magnetic field cannot be
larger than
$\sim 10^{-9}$ Gauss and that it is coherent on scales below $\sim 1$ Mpc
\cite{LSOS}. In this field the typical Larmor radius is $r_L(E)\simeq
10^2 $ Mpc $ E_{20}/ B_{-9}$, where
$E_{20}$ is the
energy of a particle in units of $10^{20}$ eV and $ B_{-9}$ is
the magnetic field in units
of $10^{-9}$ Gauss. The typical deflection angle from the
direction of the source, located at distance $d$ can be estimated
assuming that the particle makes a random walk in the magnetic field
\cite{Wax}:
\begin{equation}
\theta(E) \simeq 3.8^o \left( \frac{d}{50 Mpc} \right)^{1/2}
\left( \frac{l_{coh}}{1 Mpc} \right)^{1/2}
\left( \frac{E}{10^{20} eV} \right)^{-1}
\left( \frac{B}{10^{-9} G} \right).
\label{eq:angle}
\end{equation}
Here we use the fact that the Larmor radius of the particle, $r_L$,  is
much larger than the coherence scale of the field,  $l_{coh}$. Given the
small angle in eq. (\ref{eq:angle}), it is puzzling that the search for 
possible sources in the error boxes of the observed events found no
candidates. 

As the strength of the magnetic field increases, the Larmor radius
decreases and the hypothesis of straight line propagation between
scatterings, that eq. (\ref{eq:angle}) relies upon, breaks down. If the
LSC has fields of $10^{-8} - 10^{-7}$ Gauss as we propose here, a
diffusive approach is more appropriate for energies $\la 10^{20}$ eV.

The possibility of a diffusive motion for UHECR in the LSC was previously
advanced in \cite{WW} and \cite{Berez}. In ref. \cite{WW}, the authors
assumed the Virgo cluster to be the source of UHECRs and estimated the
luminosity of the source required to explain the observed fluxes above
$10^{19}$ eV. They assumed a phenomenological diffusion coefficient $D(E)
\propto E^{1/2}$ and estimated the effect of energy losses by introducing
a typical time for the losses to be compared with the propagation time
scale; both treatments are improved on here. In addition, at the time of
their work, estimates for Virgo magnetic fields were two orders of
magnitude lower than the present observed values
\cite{Vallee}. With Virgo magnetic fields reaching $\ga 10^{-6}$ Gauss,
UHECRs produced by sources inside Virgo (like M87) are unable to reach us,
since the diffusion timescale becomes larger than the timescale for losses
\cite{Blasi}.

In ref. \cite{Berez}, the authors study the  diffusive  propagation in
the LSC assuming a bursting source of UHECR. The resulting spectra are
characterized by a pronounced cut-off due to photopion production, such
that the events  above
$\sim 10^{20}$ eV cannot be fit. The problem at the highest energies is 
the result of considering a diffusive propagation all the way up to the
highest energies. 

Here we study the propagation in two different regimes; first at 
lower energies where diffusion works, and then at higher
energies, where particles propagate almost rectilinearly. For the
diffusive propagation,  we follow the analytical treatment as in
ref.\cite{Berez}, where energy losses are taken into account in the
proper way by solving the transport equation for cosmic rays:
\begin{equation}
\frac{\partial n}{\partial t} - div (D\nabla n) +
\frac{\partial \left[b(E) n\right]}{\partial E} = q(E,\vec{r},t) \ \ .
\label{eq:transport}
\end{equation}
\noindent
Here $n(E,r,t)$ is the density of cosmic rays with energy $E$  at a
distance
$r$ from the source  at time $t$, $b(E)=dE/dt$ is the rate of energy
losses,
$D(E)$ is the diffusion coefficient, and $q(E,\vec{r},t)$ is the density
of particles per unit time per unit energy interval injected by the
source. In general the source function is due to  a source distribution,
but here we consider a single source contribution at a time, which
corresponds to taking $q(E,\vec{r},t)=Q(E,t)\delta (\vec{r})$, where
$Q(E,t)\propto E^{-\gamma}$ is the differential spectrum of the source.
In general the source can be operating for a finite time $T$.

The solution to eq. (\ref{eq:transport}) can be easily found by
introducing the Green's function
\begin{equation}
G(\vec{r},E,t;\vec{r}_g,E_g,t_g)=
\frac{\delta(t-t_g-\tau)}{(4\pi\lambda(E,E_g))^{3/2}}
\frac{e^{-(\vec{r}-\vec{r}_g)^2/4\lambda}}{b(E)} \ \ ,
\label{eq:green}
\end{equation}
\noindent
where 
\begin{equation}
\tau(E,E_g)\equiv \int_E^{E_g} d\epsilon \frac{1}{b(\epsilon)} \ \ ,
\end{equation}
\noindent
and
\begin{equation}
\lambda(E,E_g)\equiv \int_E^{E_g} d\epsilon \frac{D(\epsilon)}
{b(\epsilon)}.
\end{equation}
For a point source the general solution of eq. (\ref{eq:transport}) 
can be written as
\begin{equation}
n(E,r,t) = \int_{-\infty}^{+\infty} dt_g \int_{0}^{E_{max}}
Q(E_g) G(r,E,t;r_g,E_g,t_g) d E_g,
\end{equation}
\noindent
which, after using eq. (\ref{eq:green}) above, becomes
\begin{equation}
n(E,r,t) = \frac{1}{b(E)} \int_{E_g(E,t-T)}^{E_g(E,t)} d E_g Q(E_g)
\frac{e^{-r^2/4\lambda(E,E_g)}}
{\left[4\pi\lambda(E,E_g)\right]^{3/2}} \ \ .
\label{eq:solution}
\end{equation}
The limits of integration $E_g(E,\tilde{t})$ are calculated integrating
the equation $dE/dt=b(E)$ back to time $\tilde{t}$ if the detected energy
is  $E$. More precisely the upper limit should be written as
$min\left[E_{max},E_g(E,t)\right]$, where $E_{max}$ is the maximum energy
that can be produced in the source.

Some comments about eq. (\ref{eq:solution}) are in order: {\it i)} if the
source is continuous or  operates on time scales larger than the typical
propagation time (which is a function of energy in the diffusive
approach), then this equation reduces to the well known equilibrium
time-independent solution; {\it ii)} the energy losses which are relevant
in the energy range above $\sim 10^{19}$ eV are due to pair production
($p+\gamma_{MW}\to e^+ + e^- + p$) off  microwave (MW) background
photons, and photopion production ($p+\gamma_{MW}\to \pi + N$, where $N$
is a proton or a neutron according to the charge sign of the pion). The
second channel becomes relevant above $\sim 3 \times 10^{19}$ eV and
determines the absorption of particles which propagate for times longer
than $\sim 10^8$ yrs. At energies below
$\sim 3\times 10^{19}$ eV, the typical timescales for energy losses (due
to pair production) are longer than the propagation time, so that energy
losses do not play an important role. In this case eq. (\ref{eq:solution})
reduces to the purely diffusive solution
\begin{equation}
n(E,r) = \frac{Q(E)}{4\pi r D(E)}\ .
\end{equation}
\noindent
This shows that the volume suppression in the
diffusive regime (without losses) goes as $1/r$ instead of $1/r^2$ and
 that the diffusive
solution results in a spectrum $n(E)\propto E^{-(\gamma+\eta)}$, if the
diffusion coefficient is $D(E)\propto E^{\eta}$.

These comments are no longer valid in the energy region where photopion
production becomes relevant. Moreover,  at  high enough energies, the
diffusion approximation breaks down and particles start to propagate in
close to straight lines. The precise energy at which this transition
between diffusion and straight line propagation occurs is difficult to
determine without a specific model for the magnetic field in the LSC.
Here we consider a transition {\it range} of energies where both the
diffusive and the straight line propagation approaches are insufficient. 
Given a specific model for the LSC magnetic field, numerical simulations
are better suited to study  the spectrum in the transition regime
\cite{sigl}. Numerical simulations are therefore complementary to the
approach given here, since the diffusion coefficient can more easily
account for the structure of  magnetic fields on scales below numerical
resolutions.

The diffusion coefficient can be calculated by
\begin{equation}
D(E)=\frac{1}{3} r_L c \frac {B^2}{\int_{1/r_L}^{\infty}
dk P(k) } \ ,
\label{eq:diffcoef}
\end{equation}
where $P(k)$ is the magnetic field power spectrum. At energies which
correspond to Larmor radii smaller than the scale of magnetic cells
(i.e., $r_L(E)\la l_{cell}$), we assume that the particles experience a
Kolmogorov spectrum of magnetic field fluctuations, $P(k)=P_0
(k/k_0)^{-5/3}$, where  $k_0\sim 1/l_{cell}$ is the small wavenumber
limit for the magnetic field and
$3 P_0 k_0/2 = B^2$. The corresponding diffusion coefficient is therefore
$D(E)\propto E^{1/3}$. For energies such that $r_L(E)\ga l_{cell}$, the
diffusion coefficient is linear in energy, $D(E)\propto E$.

 The Kolmogorov spectrum can be justified, if the magnetic field in the
LSC has structure similar to a cascade from the largest scales $l_{cell}$
to small scales. This may be the case if outflows from galaxies and
peculiar velocities of the plasma in the LSC have experienced enough
turbulent motion as in \cite{CROK} and \cite{Bierm}. However, for the
limited  range of energies where UHECRs are expected to be
extragalactic,   the resulting spectrum is not strongly dependent on the
specific choice of the power index $\eta$. 

More precisely,  we can write the diffusion coefficient as
\begin{equation}
D(E) = D_{low}(E) = 1.1\cdot 10^{34} 
\left(\frac{E}{10^{19}eV}\right)^{1/3}
\left(\frac{B}{5\cdot 10^{-8} G}\right)^{-1/3}
\left(\frac{l_{cell}}{0.5 Mpc}\right)^{2/3}~{\rm cm}^2/{\rm s}
\label{eq:dlow}
\end{equation}
\noindent
for $E\lesssim E_c=2.3\cdot 10^{19}\left(\frac{l_{cell}}{0.5 Mpc}\right)
\left(\frac{B}{5\cdot 10^{-8} G}\right)$ eV, and
\begin{equation}
D(E) = D_{high}(E) = D_{low}(E_c) \left(\frac{E}{E_c}\right)
\end{equation}
\noindent
for $E> E_c$.

We can determine the limits of the diffusive approach as follows. The
diffusion time from a distance $r$ for particles with energy $E$ is given
by $\tau_{diff}=r^2/4D(E)$, while the time required for the straight line
propagation is just $\tau_s=r/c$; the condition of diffusive propagation
is then $\tau_{diff}\gg \tau_s$, or equivalently, for fixed $r$, that
$E\ll 4\cdot 10^{20} eV (r/10~Mpc)$ for the fiducial values of magnetic
field and coherence length as in eq. (\ref{eq:dlow}). This condition can
also be interpreted as an upper limit on the diffusion coefficient:
$D(E)\ll rc/4 = 2.3\cdot 10^{35} (r/10~Mpc)~cm^2/s$, independently of the
particular energy dependence of the diffusion coefficient.

When the condition $\tau_{diff}\gg \tau_s$ is no longer fulfilled then
the propagation regime changes and cosmic rays start moving on almost
rectilinear trajectories. In this regime, we estimate  the number of
particles arriving from a source according with the following procedure:
the particles reaching us with a fixed energy $E$ must have left the
source with a {\it generation energy} $E_g$ which can be derived by
integration of the energy loss equation
$dE/dt=b(E)$ for a time given by $t=r/c+t_d$, where $t_d$ is the time
delay defined as
\begin{equation}
t_d=1.1\times 10^7 \left( \frac{E}{10^{20}eV}\right)^{-2}
\left( \frac{l_{cell}}{0.5~Mpc}\right)
r_{10}^2 \left(\frac{B}{5\cdot 10^{-8} G}\right)^2~{\rm yrs}.
\end{equation}
The time $t$ is thus the total minimum propagation time of particles
with energy $E$ from a source at distance $r$. The flux of particles
with energy $E$ is given by
\begin{equation}
\Phi(E,r)=\frac{1}{4\pi r^2} Q(E_g(E)) \frac{dE_g}{dE},
\label{eq:rp}
\end{equation}
\noindent
where $Q(E_g(E))$ is the number of particles with energy $E$ emitted back
at time $t$  with energy $E_g$, per unit energy and per unit time. For
the LSC redshift effects are not relevant and  it can be shown  that
$dE_g/dE=b(E_g)/b(E)$ \cite{BG}, which corresponds to ignoring
 evolutionary effects.

Although less probable,  time delays larger than $t_d$ are still
possible, so that a better expression for the flux should include
an average on the probability distribution of  time delays $P(t_d)$,
which is given in ref. \cite{Wax}. The flux thus becomes:
\begin{equation}
\Phi(E,r)=\frac{1}{4\pi r^2} \int_{t_d}^{\infty}
Q(E_g(t_d)) \frac{dE_g}{dE} P(t_d) dt_d,
\label{eq:rp1}
\end{equation}
\noindent
which however does not give results appreciably different from
eq. (\ref{eq:rp}).

In this paper, we assume that the source spectrum is of the form
$Q(E)=KE^{-\gamma}$, with $K$ related to the total cosmic ray luminosity
$L_p$ through $K=L_p (\gamma - 2)$, assuming $1$ GeV as the minimum
energy in the cosmic ray spectrum. An important point is that in the
limit of very high energies, eqs. (\ref{eq:rp}) and (\ref{eq:rp1}) must
recover the source spectrum. In fact, for $E\gtrsim (2-3)\times 10^{20}$
eV the time scale for energy losses is approximately constant,
$(1/E)(dE/dt)\simeq constant$, so that
$b(E)$ is linear in energy and $dE_g/dE$ turns out to be a constant. Thus
in this picture a flattening of the spectrum at high energy is a
straightforward consequence of the transition between the diffusion
and the straight line propagation regimes.

In this section, we have  identified two limits that can be used  to
understand the general features of the UHECR spectrum: at relatively low
energies, $E\la 10^{19}$ eV, where energy losses are not important, the
flux of particles detected per unit solid angle, per unit surface, per
unit energy and per unit time is is well approximated by $I(r,E)=n(E,r)
c/(4\pi) \propto Q(E)/ r D(E)$, where the factor of $4\pi$ comes from the
fact that diffusion makes the flux locally isotropic. The spectrum then
goes as
$E^{-(\gamma+\eta)}$. At very high energies the spectrum flattens to the
source spectrum, and the space dilution factor is again proportional to
$1/r^2$. In this case, the solid angle covered by a given source decreases
with energy. For a distribution of sources in the LSC, the flux per unit
solid angle can be estimated by $I(E,r)\approx \Phi(E,r)/(2\pi)$, since
the LSC covers about half an hemisphere in the sky. In the following
section, we find the spectrum of UHECRs for specific source parameters.

\section{Sources of UHECRs in the Local Supercluster}

In Figs. 1 and 2, we plot the spectrum as a function of energy for a
continuous source located between 10 and 17 Mpc away from us. We vary the
LSC average magnetic field between $5 \times 10^{-8}$ and $10^{-7}$
Gauss. The required luminosity for the source varies according to the
spectral index, such that for $\eta = 2.1$, $L_p \simeq 10^{43}$ erg/s is
sufficient while for $\eta = 2.4$, $L_p \ga 10^{45}$ erg/s. In Fig. 1, we
assume that the maximum energy reached at the source is $E_{max} =
10^{21}$ eV, while in Fig. 2, $E_{max} = 10^{22}$ eV. 

Contrasting our calculations with the data from ref. \cite{data}, we see
that the solid and dotted lines can reproduce the spectrum reasonably well
within the observed uncertainties.  The flattening of the spectrum at
higher energies  is  clear  and the flux level is compatible with the 
data. These two cases correspond to sources at 13 Mpc (solid) and 10 Mpc
(dotted) with fields between 5 and 10 $\times 10^{-8}$ Gauss,
respectively. 

These choices for the magnetic field of the LSC and distance to the
source give good fits for the following reason. The diffusive regime is a
good approximation for energies below $E_0= 8 \times 10^{19} B_{-8}
r_{10}$ eV, for
$l_{cell}\simeq 0.5$ Mpc. Therefore,  if we want to reproduce the
flattening of the spectrum at  $\sim 1-3 \times10^{20}$ eV, then we need
$B_{-8} r_{10}$ of a few.  For magnetic fields in the range that we
propose here, $B_{-8} \sim 5 - 10$, the sources have to be relatively
nearby at about 10 Mpc.  Increasing the distance to the source causes the
flattening point to move to higher energies and the flux of cosmic rays
to decrease drastically, with the appearance of a pronounced cut-off. In
other words, a magnetized LSC brings the GZK cutoff to much shorter
distances. 

In addition, bursting sources cannot fit the spectrum if the LSC has
fields as strong as we suggest here. The need to fit both the energies
below the GZK cutoff as well as those above, constrains the sources to be
continuous over a long enough period ($\ga $ Gyr) for the observed flux to
reach the a steady-state signature.

If  UHECRs are accelerated in active galaxies, there are a number of
possible nearby sources within a 20 Mpc radius of our Galaxy. For
example, NGC 3031, M82, NGC 3115, and Cen A (NGC 5128) are all below 4
Mpcs away; NGC 4051 is about 10 Mpc away; and NGC 4151, NGC 3227, NGC
1068, NGC 1566, and M87 are between 15 and 18 Mpc away. If the source is
very close, than the deflection angles become small at the highest
energies, and a counterpart to the source should be identifiable. The
stronger the LSC field, the larger the typical deflection angle for the
sources. Numerical simulations indicate that the deflection angle can be
of the order of 10$^o$ for our choices of parameters \cite{sigl}.

\section{Conclusion}

In conclusion, we have shown that a  sufficiently magnetized Local
Supercluster can explain the spectrum  of ultra-high energy cosmic rays.
The spectrum of extragalactic cosmic rays with energies below $\sim
10^{19}$ eV may be due to the diffusive propagation in the Local
Supercluster with fields of  $\sim 5 - 10 \times 10^{-8}$ Gauss. Above 
$\sim 10^{20}$ eV, cosmic rays propagate in an almost rectilinear way
which is evidenced by the change in shape of the spectrum at the highest
energies.  The  transition from diffusive to straight line propagation of
UHECRs in the LSC gives a good fit to the spectral changes observed with
current experiments as long as at least one  source be located relatively
nearby at $\sim 10$ Mpc away from our Galaxy.

Future studies of the magnetic field structure of the LSC as well as the
detection of UHECRs at the highest energies will help
determine the sources of UHECRs. Another consequence of our model is that
the sources are relatively nearby, thus, the precise structure at the edge of
the LSC is not as important as the ``local'' 10 to 15 Mpc radius around us.
Finally,  we expect a significant North-South asymmetry in the UHECR data at
the highest energies, as the LSC is more prominent in the Northern
Hemisphere. This asymmetry could be detected by future experiments that
cover both hemispheres such as the proposed Auger Project \cite{Auger}.  

In addition, with respect to the angular distribution of the observed
events,  the lack of counterpart identifications  for energies below
$10^{20}$ eV is a natural consequence of a close to diffusive propagation
in the LSC. We predict that counterparts will be more easily identified  
at energies above the limit of detectability of current experiments. 

These findings show that a simple model for UHECRs based on one or more
nearby continuos sources can explain the presently available data on
UHECRs, if the LSC has relatively strong magnetic fields ($\sim
10^{-8}-10^{-7}$ Gauss).  Such fields are allowed by current limits on
extragalactic magnetic fields and could indicate the presence of magnetic
fields on cosmological scales. 

\acknowledgements

We gratefully acknowledge many illuminating discussions with
G. Sigl, M. Lemoine,  V.S. Berezinskii, R. Rosner, P. Kronberg, and P.
Biermann. P. B. was supported by a Istituto Nazionale di Fisica Nucleare
fellowship at the University of Chicago. A. V. O. was supported in part by
the DOE through grant DE-FG0291 ER40606, and by the NSF through grant
AST-94-20759.

\begin{figure}
\noindent{ FIG. 1. Flux versus Energy assuming a maximum energy at the
source of $E_{max} = 10^{21}$ eV. The  solid line corresponds to $r=$13 Mpc,
$B = 5
\times 10^{-8}$ Gauss,
$\gamma = 2.1$, and
$L_p= 2.2 \times 10^{43}$ erg/s; dotted line is for  $r=$10 Mpc, $B =  
10^{-7}$ Gauss, $\gamma = 2.1$, and
$L_p= 10^{43}$ erg/s; dashed line is for  $r=$10 Mpc, $B = 
10^{-7}$ Gauss, $\gamma = 2.4$, and $L_p= 3.2 \times 10^{45}$ erg/s;  and
dashed-dotted line corresponds to $r=$17 Mpc, $B = 5 \times
 10^{-8}$ Gauss, $\gamma = 2.1$, and
$L_p= 3.3 \times 10^{43}$ erg/s. The data is from ref.\cite{data}.}
\end{figure}

\begin{figure}
\noindent{ FIG. 2. Flux versus Energy assuming a maximum energy at the
source of $E_{max} = 10^{22}$ eV. The  solid line corresponds to $r=$13 Mpc,
$B = 5
\times 10^{-8}$ Gauss,
$\gamma = 2.1$, and
$L_p= 2.2 \times 10^{43}$ erg/s; dotted line is for  $r=$10 Mpc, $B =  
10^{-7}$ Gauss, $\gamma = 2.1$, and
$L_p= 10^{43}$ erg/s; dashed line is for  $r=$10 Mpc, $B = 
10^{-7}$ Gauss, $\gamma = 2.4$, and $L_p= 3.2 \times 10^{45}$ erg/s;  and
dashed-dotted line corresponds to $r=$17 Mpc, $B = 5 \times
 10^{-8}$ Gauss, $\gamma = 2.1$, and
$L_p= 3.3 \times 10^{43}$ erg/s. The data is from ref.\cite{data}.}
\end{figure}

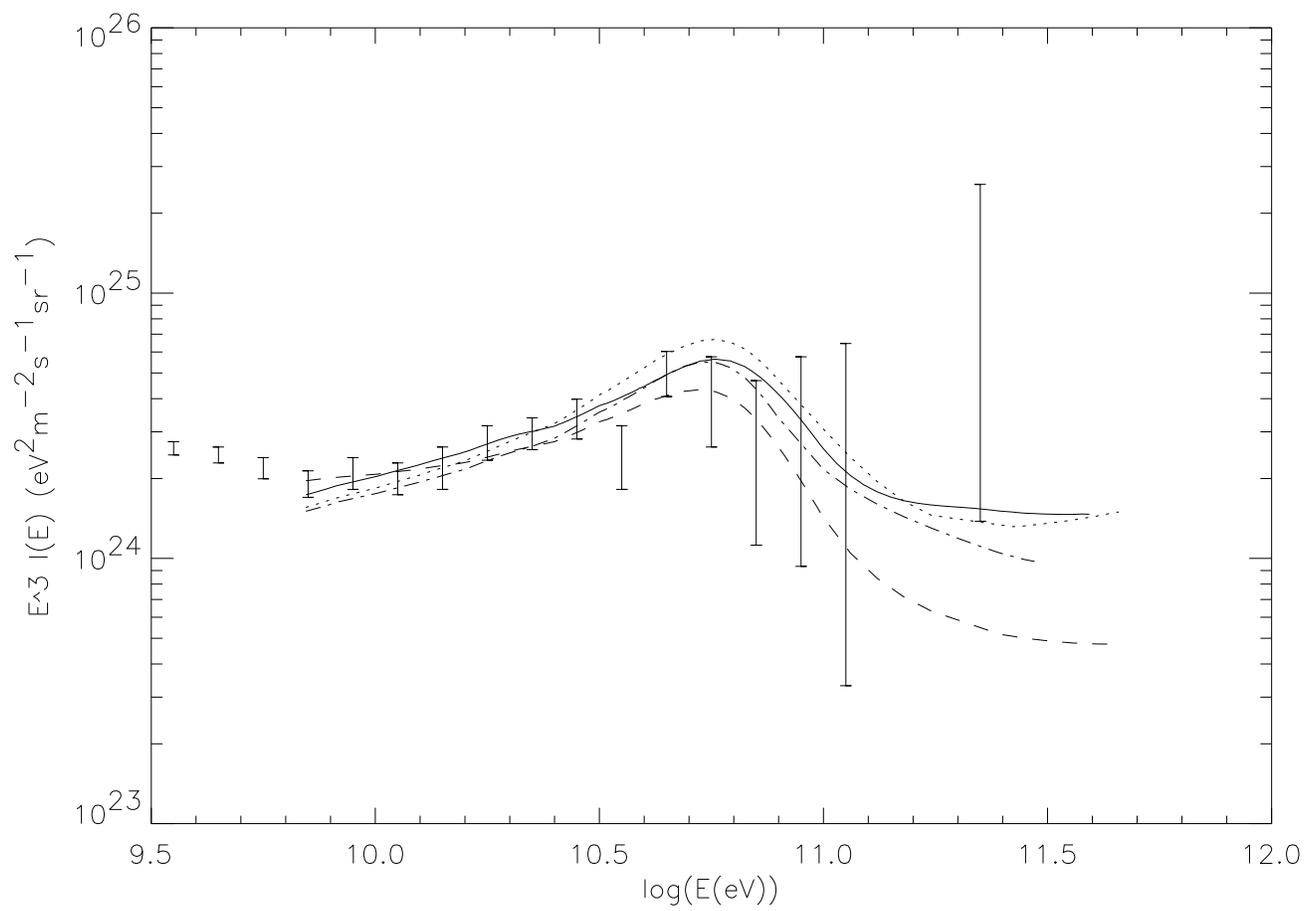
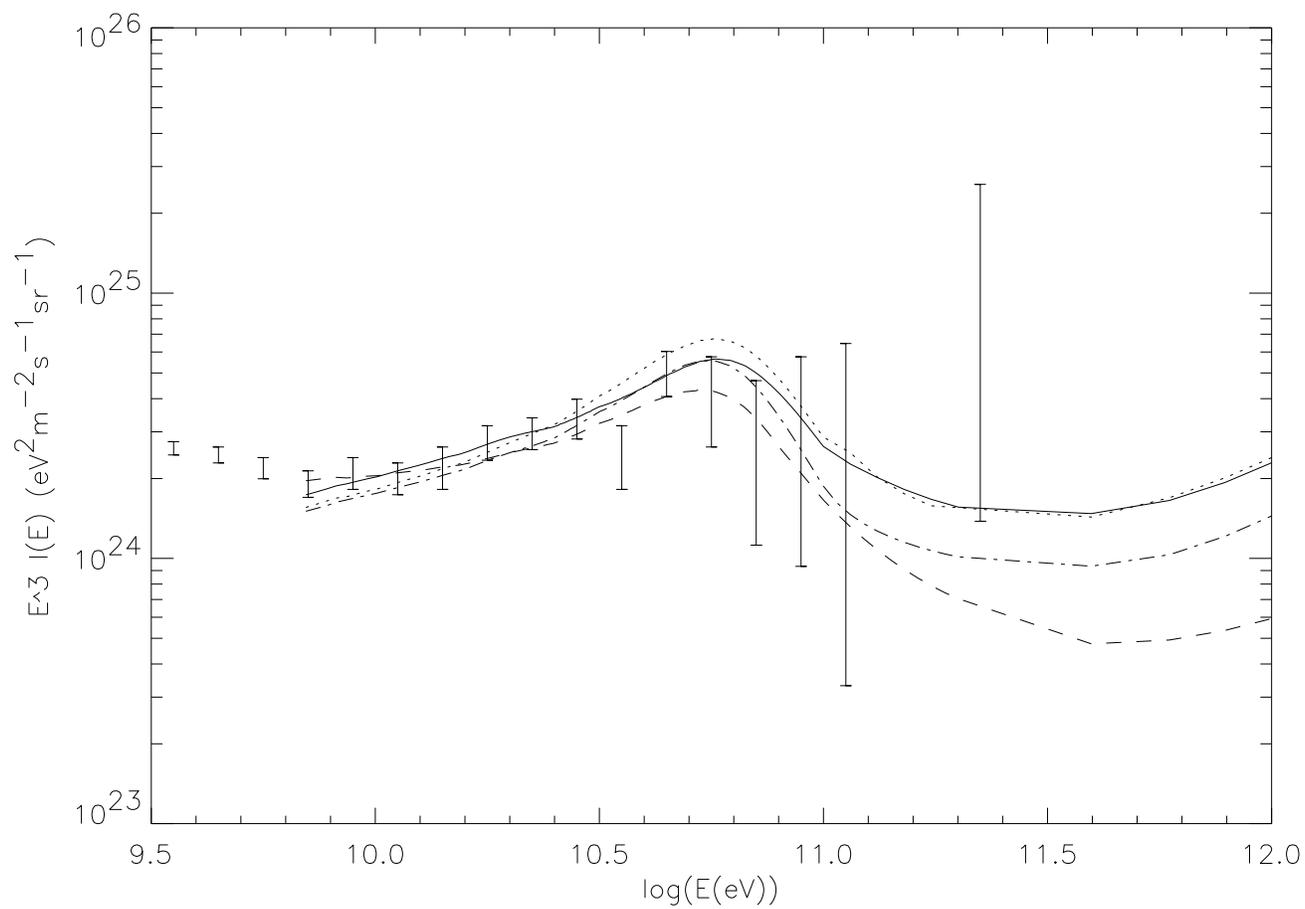
\end{document}